\documentclass[floatfix,twocolumn,aps,showpacs,prl,superscriptaddress,longbibliography]{revtex4-1}

\usepackage{graphicx}
\usepackage{multirow,booktabs}

\usepackage{amsmath,amssymb,latexsym,MnSymbol}
\usepackage[colorlinks=true, urlcolor=blue,linkcolor=blue,citecolor=blue]{hyperref}

\makeatletter
\let\cat@comma@active\@empty
\makeatother

\usepackage{lineno}
\usepackage{verbatim}

\begin{document}

\title{Long-range correlations and fractal dynamics in \textit{C. elegans}: changes with aging and stress}

\author{Luiz G. A. Alves}\email{lgaalves@dfi.uem.br}
\affiliation{Department of Chemical and Biological Engineering, Northwestern University, Evanston, IL 60208, United States of America}
\affiliation{Department of Physics, State University of Maring\'a, Maring\'a, PR 87020-900, Brazil}
\affiliation{National Institute of Science and Technology for Complex Systems, CNPq, Rio de Janeiro, RJ 22290-180, Brazil}
\author{Peter B. Winter}\email{peterwinter@u.northwestern.edu}
\affiliation{Department of Chemical and Biological Engineering, Northwestern University, Evanston, IL 60208, United States of America}
\author{Leonardo N. Ferreira}
\affiliation{Department of Chemical and Biological Engineering, Northwestern University, Evanston, IL 60208, United States of America}
\affiliation{Institute of Mathematics and Computer Science, University of S\~ao Paulo, S\~ao Carlos, SP 13566-590, Brazil}
\author{Ren\'ee M. Brielmann}
\affiliation{Department of Molecular Biosciences, Northwestern University, Evanston, IL 60208, United States of America}
\author{Richard I. Morimoto}
\affiliation{Department of Molecular Biosciences, Northwestern University, Evanston, IL 60208, United States of America}
\author{Lu\'is A. N. Amaral}\email{amaral@northwestern.edu}
\affiliation{Department of Chemical and Biological Engineering, Northwestern University, Evanston, IL 60208, United States of America}
\affiliation{Department of Physics and Astronomy, Northwestern University, Evanston, IL 60208, United States of America}

\begin{abstract}
{Reduced motor control is one of the most frequent} features associated with aging and disease. Nonlinear and fractal analyses have proved to be useful in investigating human physiological alterations with age and disease. Similar findings have not been established for any of the model organisms typically studied by biologists, though. If the physiology of a simpler model organism displays the same characteristics, this fact would open a new research window on the control mechanisms that organisms use to regulate physiological processes during aging and stress. Here, we use a recently introduced animal tracking technology to simultaneously follow tens of \textit{Caenorhabdits elegans} for several hours and use tools from fractal physiology to quantitatively evaluate the effects of aging and temperature stress on nematode motility. Similarly to human physiological signals, scaling analysis reveals long-range correlations in numerous motility variables, fractal properties in behavioral shifts, and fluctuation dynamics over a wide range of timescales. These properties change as a result of a superposition of age and stress-related adaptive mechanisms that regulate motility. 

\end{abstract}


\maketitle
Fractal-like {fluctuations} are a hallmark of healthy physiological systems such as heart rate~\cite{Goldberger2000,Walleczek2006}, neural spiking~\cite{Goldberger2000,Walleczek2006}, and gait dynamics of humans~\cite{Hausdorff2007}. The widespread prevalence of fractal-like dynamics in physiological processes refuted classical theories of physiological control, which assumed that health is maintained through strict homeostasis and that fluctuations away from homeostasis should be uncorrelated. Instead, physiological signals show self-similar patterns across multiple scales and exhibit long-range correlations in their fluctuations.

Fractal-like patterns are {also} widespread in animal behavior such as the timing of specific movements and diffusive patterns in the paths of animals moving through their environment. For example, it has been argued that L\'evy flights are an optimal strategy for landscape exploration in the search for food, sexual partners, and so on~\cite{Mandelbrot1983}. L\'evy flights have been observed {\bf in the} foraging behavior of ants~\cite{Stanley1985}, albatrosses~\cite{Viswanathan1996}, monkeys~\cite{Ramos-Fernandez2004}, sharks, bony fishes, sea turtles, and penguins~\cite{Sims2008}. Fractal patterns have also been observed in the timing of specific behaviors, such as feeding, sexual, social, and vigilant behavior in Spanish ibexes~\cite{Alados1996}, fathead minnows~\cite{Alados1999}, wild chimpanzees~\cite{Rutherford2003}, and domestic hens~\cite{Rutherford2003}, respectively.

Here, we examine the behaviors of one of the simplest multicellular model organisms, \textit{Caenorhabditis elegans}, and find that it displays fractal-like movement dynamics. \textit{C. elegans} is a prominent model organism {in molecular biology} because of its simple body structure and a fixed cell lineage containing 302 neurons from a total of 959 somatic cells. Despite its relative simplicity, the nematode shares many biological characteristics with more complex organisms such as humans.
They have an organ system which includes a digestive system, a nervous system, gonads, and muscles~\cite{McMullen2012, Eisenmann2005}. They have a well-characterized life-cycle involving development,
reproduction, and aging~\cite{Wolkow2012,Bansal2015}. Despite their small genome size ($\sim$100 Megabase versus 3.6 Gigabase for humans), nearly 40\% of {its} genes are human homologs~\cite{Consortium1998}, and the majority of human disease genes and disease pathways are present in this nematode~\cite{Lai2000,Kaletta2006}. These commonalities make \textit{C. elegans} an ideal model organism for {experimentally} studying health and behavior.

\begin{figure}[!ht]
\begin{center}
\includegraphics[width=0.45\textwidth]{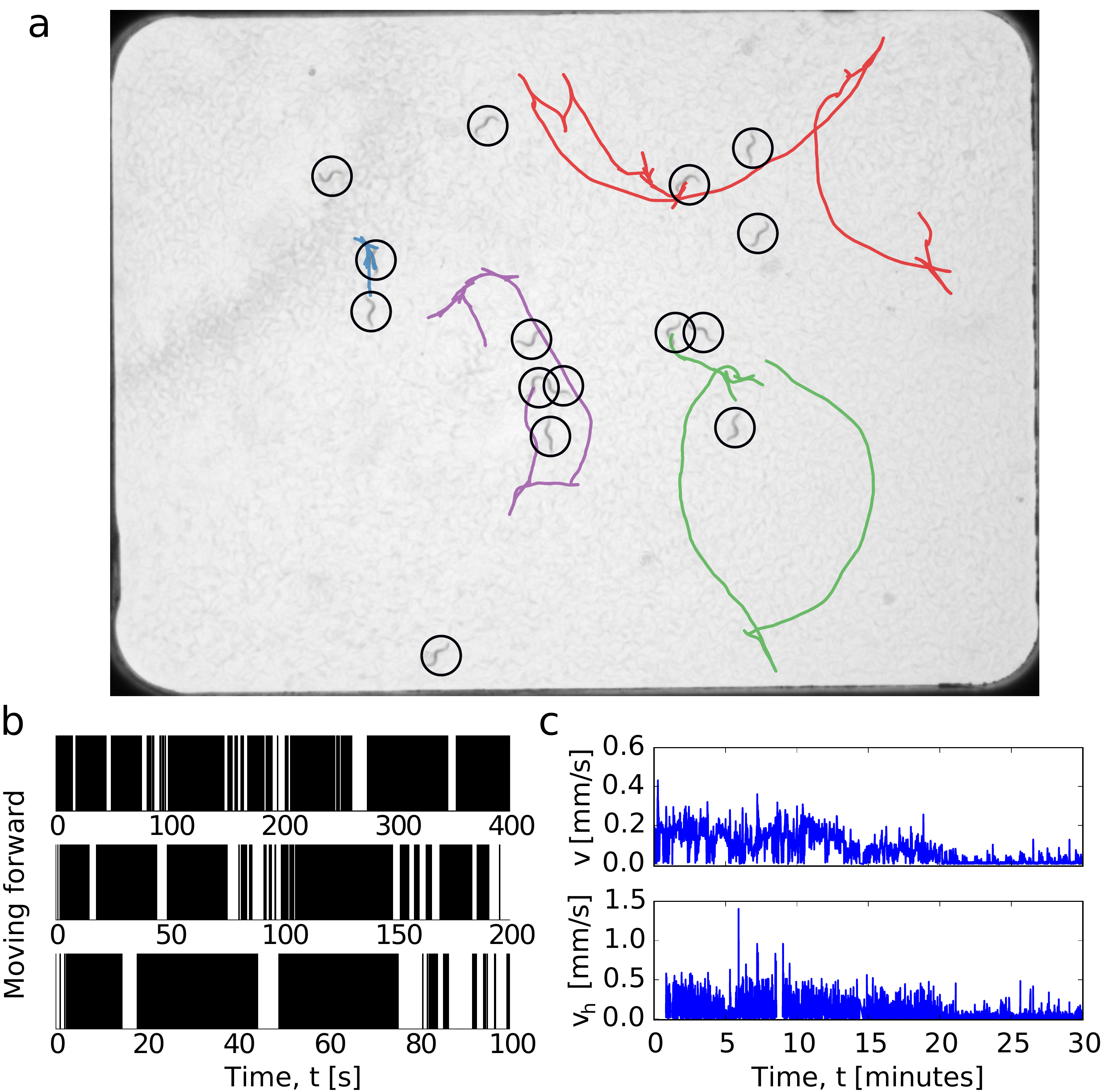} 
\end{center}
\caption{{Self-similarity of \textit{C. elegans} motility}. (a) Our experimental system enables us to track ten to fifteen worms (at a time) that are confined within a cooper enclosure $2.5$~cm $\times 1.5$~cm, the equivalent of a basketball court for humans. We plot the trajectories of four worms during a ten minute period. Notice that the variability in behaviors across individuals. See Supplemental Material for a video from a single worm~\cite{video}. (b) Intermittent behaviors, such as ``moving forward" display a Cantor dust-like behavior, indicating fractality. {The black vertical bars represent periods of forward motion and the white ones the absence of this behavior}. (c) Centroid speed $v$ and head speed $v_h$ time series {of a single worm} exhibit fluctuations across a broad range of timescales.}
\label{fig:1}
\end{figure}

In fact, many aspects of \textit{C. elegans} behavior have already been linked to specific biological processes. Aspects of \textit{C. elegans} motility have been linked to specific neurons~\cite{Haspel2010}, genes~\cite{Brenner1974}, and environmental stimuli~\cite{Luo2014}. Many behavioral metrics have been studied for \textit{C. elegans}, including speed~\cite{Ramot2008,Hahm2015}, body posture~\cite{Brown2013}, frequency of particular actions~\cite{Herndon2002}, and the configuration of the worm's body over time~\cite{Stephens2008}. Despite having a nearly isogenic background, individual nematodes raised under the same conditions can have a high degree of individual variability in movement-related behaviors~\cite{Winter2016}. Furthermore, even individual \textit{C. elegans} can show highly variable behavior {when observed} for time periods longer than a few minutes (Fig.~\ref{fig:1}).

In order to create a sufficient number of multi-hour time series tracking the behavior of individual animals, we use the Multi-Worm Tracker's real-time data acquisition~\cite{Swierczek2011} software and correct imaging and worm identity errors after acquisition using the Worm Analysis for Live Detailed Observation (WALDO)~\cite{Winter2016} software. Our experimental and software infrastructure, allow us to track tens of animals at a time for multiple hours while still maintaining the identities of individual animals. 

The methods used to acquire all motility data for this paper were previously described in detail by Winter \emph{et al}~\cite{Winter2016}. We used Wild-type Bristol isolate of \textit{Caenorhabditis elegans} (N2) from the Caenorhabditis Genomic Center (CGC) for all experiments. Standard methods were used for culturing and observing \emph{C. elegans}~\cite{Brenner1974}. {Nematodes were age-synchronized via egg-laying and grown to adulthood at $20^{\circ}$C on 60 mm nematode growth medium (NGM) plates seeded with 200~$\mu$L of \emph{Escherichia coli} OP50 strain. The plates were swirled until they reached a uniform distribution of food across their surfaces}. Ten to fifteen animals were placed on a 60 mm NGM plate inside a copper frame with $2.5 \times 1.5$ cm interior dimensions. All motility assays were performed inside of a Percival I-36NL C8 incubator to ensure a nearly constant environmental temperature.

\begin{figure}[!ht]
\begin{center} 

\includegraphics[width=0.45\textwidth]{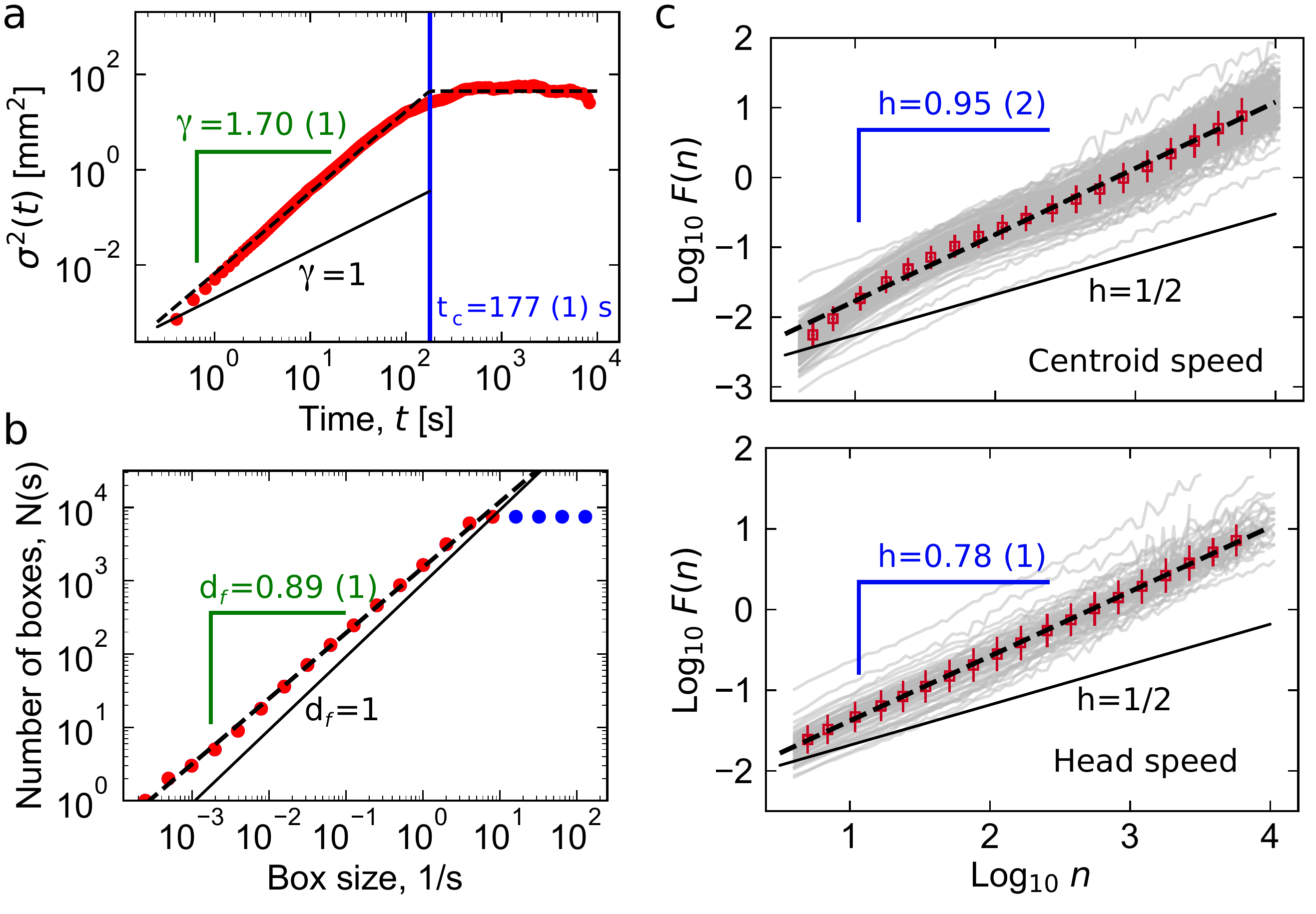} 
\end{center}
\caption{{Fractality of \textit{C. elegans} motility.} (a) Determination of diffusive behavior of worms at day one of adulthood. Red dots are the data, and the black dashed line is a fit to Eq. (\ref{eq:1}) {and the continuous line represents a random diffusion}. $\gamma$ is significantly larger than 1, indicating that movement of the worm is not random.  (b) Determination of fractal dimension of intermittent behavior ``moving forward" using the box-counting method. $d_f$ is smaller than 1, suggesting that forward motion is not the default behavior, that is, that there are periods of all lengths in between consecutive periods of forward motion. {The continuous line represents a time series where only exist forward motion}. (c) The determinant of long-range correlations in velocities time series. Each gray line is the fluctuation $\log_{10} F(n)$ as a function of the scale $\log_{10} n$ for a centroid and head speed time series. The square red dots represent binned averages over all curves and error bars are standard deviations. The dashed black line is OLS fit to the averages. {The continuous line represents a random process with $h=1/2$}. The Hurst exponent $h>1$ indicates that velocities have long-range persistent correlations. The numbers between parentheses in all plots are the standard error in the last digit. } 
\label{fig:2}

\end{figure}

The time series shown in Fig.~\ref{fig:1} display irregular patterns that are linked to how the organism processes information about its internal state and the chemical and mechanical cues from its surroundings. {For example, forward or backward motion has been related to the activity of specific groups of neurons during foraging behavior~\cite{roberts2016stochastic}}. We focus on three types of analyses commonly used to detect fractal behavior: Mean Square Displacement (MSD), fractal dimension, and long-range correlations~\cite{Metzler2014,Ribeiro2014,Reverey2015,Alves2016,Falconer1990,Peng1994,Goldberger2000,Goldberger2002}. 
By implementing all of these approaches, we assess whether individual worms change their position, and regulate movement in a manner consistent with fractal physiology (Fig.~\ref{fig:2}).

Mean Square Displacement (MSD) quantifies how an animal moves from its current position. We have considered the positions time series $\vec{r_i}(t)$ to measure the time dependence of the variance of the radial position, this is, $\sigma^2 (t) =\langle [\vec{r_i}(t)-\langle \vec{r_i}(t)\rangle]^2\rangle$, where $\langle \vec{r_i}(t)\rangle$ is the average radial position over all tracks $i$ at time $t$. For a random process (Brownian motion) the variance of the position of an individual increases linearly with time. More generally, the variance increases with time in a power-law fashion~\cite{Metzler2014,Ribeiro2014,Reverey2015,Alves2016}, $\sigma^2 (t) \sim t^\gamma$, where $0<\gamma<1$ corresponds to subdiffusion, $1<\gamma<2$ to superdiffusion, $\gamma=2$ to a ballistic diffusion, and $\gamma=1$ is the memory-less Brownian diffusion regime. For L\'evy flights in a bounded space, the variance can be modeled as a power-law that saturates for long times~\cite{Vahabi2013}. Mathematically, this can be written as, 
\begin{eqnarray}
\sigma^2 (t)= 
   \left\{
\begin{array}{ll}
      D\, t\,^\gamma\, & t<t_c\\
      C & t>t_c\\
\end{array} 
\right. 
,
\label{eq:1}
\end{eqnarray}
where $\gamma$ is the diffusion exponent, $D$ is a constant related to the diffusion coefficient, $t_c$ is the time need to reach the boundaries, and $C$ is constant arising from {the} confinement within a bounded area.

The results of Fig.~\ref{fig:2}(a) demonstrate that the exponent $\gamma$ is significantly different from 1 for one-day-old worms. {Superdiffusive behavior was also observed in worms recorded on a different condition, where no food were present during the data acquisition~\cite{stephens2010modes,salvador2014mechanistic}}. Our experimental data enables us to identify the power-law superdiffusive behavior ($\gamma>1$) and saturation regime of the variance for $t>100$ s consistent with the theoretical predictions for L\'evy flights in a bounded area~\cite{Vahabi2013}. We can also observe that there is a transient regime  where the data falls below the adjusted line, suggesting that $\gamma \sim 2$ would be a better fit to data for $t<10$s and that there is a transient period characterized by ballistic motion for short time scales. The ballistic behavior was also found for assays with worms  recorded on no food~\cite{stephens2010modes}.

Box-counting fractal dimensions are used to quantify the fractal nature of intermittent behaviors. \textit{C. elegans} engages in several types of intermittent behavior, such as forward and backward motion, reorientation, and coiling; Fig.~\ref{fig:1}(b). The time series of these events display a fractal geometry, that is, the structure of the signals looks similar at different timescales. To quantitatively evaluate the fractality of these signals, we use the box-counting method~\cite{Falconer1990} to calculate the fractal dimension of the intermittent behavior for every nematode. Specifically, we count the number $N(s)$ of boxes of size $s$ containing at least one non-null value. For a fractal object, $N(s) \sim s^{-d_f}$, where $d_f$ is the Hausdorff fractal dimension of the object~\cite{Falconer1990}.

For time series, the fractal dimension must be confined between $d_f=0$, when the behavior is practically absent, and $d_f=1$, when the behavior occurs with a uniform probability across time. In Fig.~\ref{fig:2}(b) we show a plot of the number of box $N(s)$ versus $1/s$ for a single worm. The fractal dimension exponent $d_f<1$ is a consequence of the unpredictability of the worm's behavior and how it reacts to cues in the environment, such as food or the concentration of excreted substances. {Indeed, it has been shown that the ability of changing behavior accordingly to external stimulus can be crucial for organism survival~\cite{roberts2016stochastic}.} The fact that we find $d_f<1$ for forward motion implies that forward motion is not the default behavior, {the worm needs to alternate the states of motion between the different movements in order to achieve an optimal search strategy.}

We next use Detrended Fluctuation Analysis (DFA) to quantify long-range correlations in the fluctuations of signals~\cite{Peng1994,Goldberger2000}. This methodology can be implemented using the following steps: \textit{i)} integrate the time series and divide it into boxes of equal length $n$; \textit{ii)} for each segment, a local polynomial trend is calculated and subtracted from the integrated profile (here we have used a linear function, but higher orders do not change our results); \textit{iii)} for a given box size $n$, calculate the root-mean-square fluctuation $F(n)$; \textit{iv)} repeat this procedure for all timescales $n$. Typically, the fluctuation function has a power-law dependence on the observation timescale $n$, $F(n)\sim n^h$. The parameter $h$ (Hurst exponent) is a scaling exponent that describes the self-similarity in the fluctuation at different timescales and is related to the decay of autocorrelation in the time series. If $h=1/2$, the time series has, at most, short-range correlations. Long-range correlations are present if $h \neq 1/2$. A $h<1/2$ signals anti-persistent changes and a $h>1/2$ signals persistent changes. 

DFA shows that both centroid and head speed time series display long-range correlations and present persistence in their velocity fluctuations for worms on the first day of adulthood. The behavior of the fluctuation function $\log_{10} F(n)$ as a function of the scale $\log_{10} n$ for the centroid speed time series $v(t)$ and head speed $v_h(t)$ of all worms are shown in Fig.~\ref{fig:2}(c). The power-law trend is clear for all individuals.

It is striking that a simple organism such as \textit{C. elegans} can display a behavior of a complexity similar to that found {for} human physiology. These findings open a new window for studying the effects of aging and stress on health, because of the shorter lives, less restrictive experimentation constraints on {invertebrate} testing and the similarities between many fundamental cellular structures and biological characteristics of \textit{C. elegans} and humans.

\begin{figure*}[!ht]
\begin{center}
\includegraphics[width=1\textwidth]{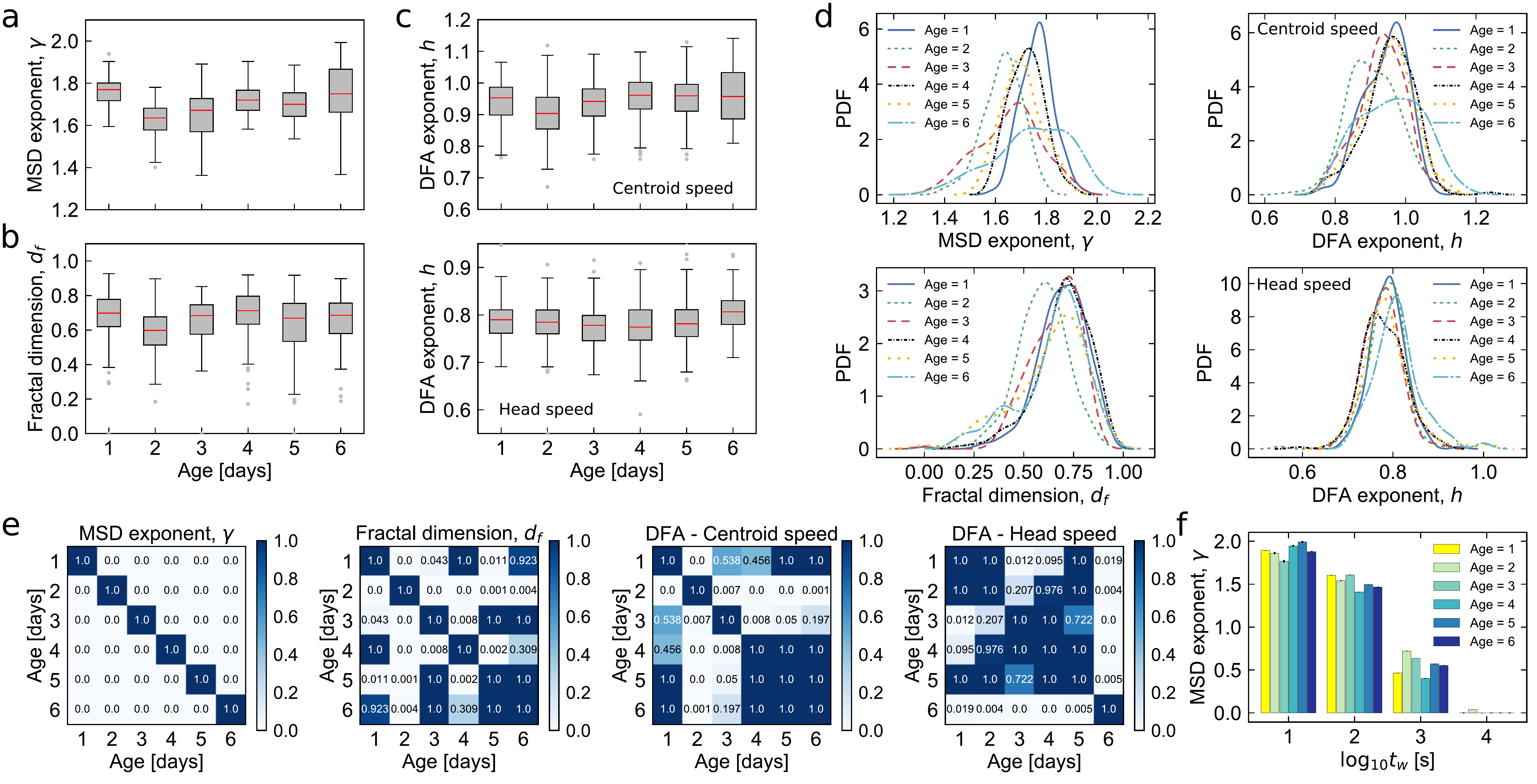}
\end{center}
\caption{{Changes in the scaling exponents with aging.} (a) {MSD exponent $\gamma$ as a function of age for 100 samples via bootstrapping}. (b) Fractal dimension $d_f$ as a function of age. (c) Long-range persistent correlations are pervasive at all ages despite changes in Hurst exponent for centroid speed and head speed time series. {(d) Probability distribution function (PDF) of the exponents calculated using kernel density estimation. The Kolomogorov-Smirnov test rejects the normal hypothesis at $95\%$ of confidence for all exponents, except for $\gamma$ at ages in the range 1 to 5.} (e) Matrices of the $p-$values resulting from the multiple comparison to test the null hypothesis that the two samples come from the same population via Mann-Whitney test with Bonferroni corrections. A $p-$value$<0.05/15$ means that the populations are distinct. {(f) MSD exponent $\gamma$ as a function of age for different time ranges $t_{w-1}<t<t_w$. The bars are the diffusion exponents $\gamma$ and the small error bars stand for the fitting standard error}. {In all box-plots, the red line represents the median, the middle ``box'' represents the middle 50\%, the upper and lower whiskers bars are the most extreme non-outlier data points, and dots are the outliers.}} 
\label{fig:3}
\end{figure*}

We know that aging and disease can drastically alter the fractal characteristics of signals from human physiology. We next test whether this is also true for \textit{C. elegans}. To explore how aging affects the dynamics of worm physiology, we repeat the previous analysis for worms of different ages (Fig.~\ref{fig:3}). 

{Using MSD, we observe the prevalence of super-diffusive behavior across all ages, but with statistically significant differences across ages. We show our estimates of $\gamma$ obtained via bootstrapping in Fig.~\ref{fig:3}(a). The distribution of exponents for each age is shown in Fig.~\ref{fig:3}(d) and the $p-$values for the Mann-Whitney test with corrections for multiples comparisons are shown in Fig.~\ref{fig:3}(e).}

As we mentioned previously, there is a transient ballistic regime for short times and because we are trying to minimize the error when fitting the data, this could leads to a diffusion exponent that do not represent well the differences across ages. To overcome this, we have calculated the MSD exponent for intervals in the range $t_{w-1}<t<t_w$ with $w=[1,4]$. Thus, in Fig.~\ref{fig:3}f we can identify three regimes: ballistic diffusion ($\gamma=2$) for $t<10$s, superdiffusion ($\gamma>1$) for $10$s $\,<t<100$s, and the saturation regime ($\gamma \approx 0$) for $t>100$s. By comparing the exponents in the region of interest (superdiffusion regime) we can see the similar pattern to what was found in Fig.~\ref{fig:3}(a).

The fractal dimension characterizing forward motion also changes with age. In Fig.~\ref{fig:3}(b), we show the statistical significance of the differences between ages, where the differences are indicated by the $p-$values for the Mann-Whitney test. We can observe that day two is slightly smaller than the other days (excluding day 5 and 6), and this could be related to egg-laying since almost 50\% of the eggs are laid in this day~\cite{McMullen2012}. {Egg-laying is known to affect movement of \textit{C. elegans}. For instance, prior to an egg-laying event, there is a transient velocity increasing and reversals movement are inhibited during egg-laying~\cite{hardaker2001serotonin}. During egg-laying, the worm stays in a state of no movement~\cite{maertens1975observations}, what could directly change the fractal exponents at this period.} Indeed, a decrease in the fractal dimension from binary behavioral time series were also observed during pregnancy for Spanish ibexes~\cite{Alados1996}. 

The DFA correlation exponent for each track -- grey lines in Fig.~\ref{fig:2}(c) -- showed a prevalence of long-range persistent correlations in the time series of centroid speed and head speed across ages, as shown in Fig.~\ref{fig:3}(c). Although the exponents are almost the same for all ages, there are some significant differences, as indicated by the $p-$values for the Mann-Whitney test. Notice that, in contrast with the differences found in human physiology where there are alterations on the DFA exponent for sick people, here we have statistically significant differences for healthy worms that only differs by their ages; Fig.~\ref{fig:3}(e).

Our results show that the fractal properties of worm motility depend on its age and life-stage. Age-related changes such as egg-laying, seeking mates or food, deterioration of organs and tissues (neuronal and muscle system) can be related to changes in the diffusion exponent $\gamma$, fractal dimension $d_f$ and Hurst correlation exponent $h$. The superpositions of these effects are manifested as small (but statically significant) changes in the exponent values. While the measured changes in exponent values appear to be quite small, one should note that the measured changes in exponent values for human heart rate variability were obtained comparing records for healthy individuals with records obtained for patients suffering from congestive heart failure, a very serious heart condition that is frequently fatal~\cite{Goldberger2000}. In contrast, our comparisons are performed for the human equivalent of a 15 yr old and a 40 yr old.

\begin{figure*}[!ht]
\includegraphics[width=1\textwidth]{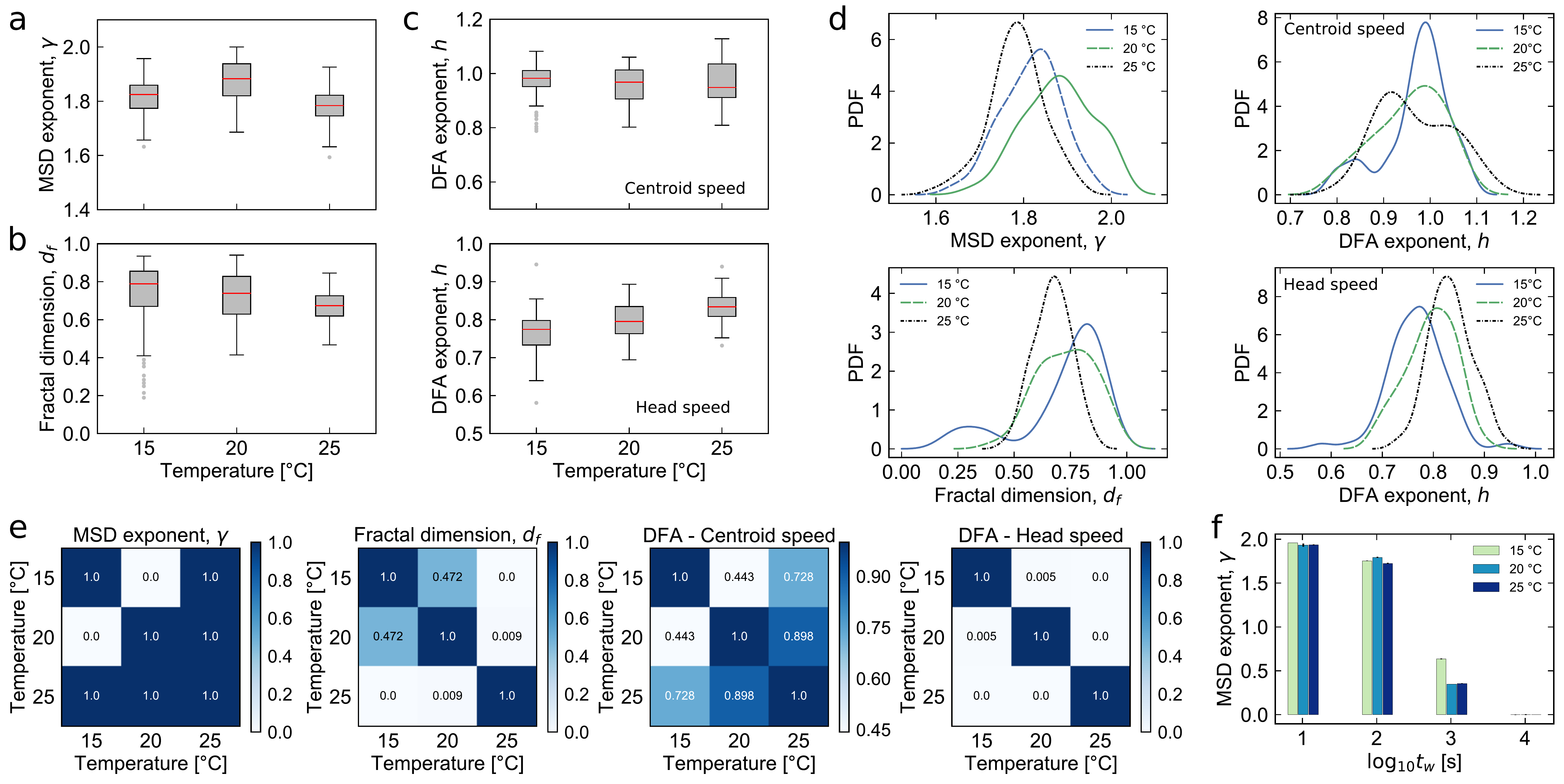}
\caption{Changes in the scaling exponents with stress. (a) {MSD exponent $\gamma$ as a function of temperature for 100 samples via bootstrapping}. (b) Fractal dimension $d_f$ as a function of age. (c) Long-range persistent correlations are pervasive at all ages despite changes in Hurst exponent for centroid speed and head speed time series. (d) Probability distribution function (PDF) of the exponents calculated via kernel density estimation. The Kolomogorov-Smirnov test rejects the normal hypothesis at $95\%$ of confidence, for  all exponents distributions, except by the distribution of MSD exponents at the temperatures $15^{\circ}$C and $25^{\circ}$C. (e) Matrices of the $p-$values resulting from the multiple comparison to test the null hypothesis that the two samples come from the same population via Mann-Whitney test with Bonferroni corrections. A $p-$value$<0.05/3$ means that the populations are distinct. (f) MSD exponent $\gamma$ as a function of temperature for different time ranges $t_{w-1}<t<t_w$. The bars are the diffusion exponents $\gamma$ and the tiny error bars stand for the fitting standard error. In all box-plots, the red line represents the median, the middle ``box'' represents the middle 50\%, the upper and lower whiskers bars are the most extreme non-outlier data points, and dots are the outliers.}
\label{fig:4}
\end{figure*}

Like aging, stress can change the fractal properties of physiological systems. {Previous works have shown that worms can change behavior according to the environmental temperature~\cite{yamada2003distribution,Luo2014,parida2014effect}. \textit{C. elegans} assays are performed at three growth temperature: $15^{\circ}$C, $20^{\circ}$C, and $25^{\circ}$C~\cite{gouvea2015experience}. The stress caused by the variation of temperature at both extremes of this range declines fecundity~\cite{byerly1976life,hirsh1976development}, can change directionality of movement~\cite{yamada2003distribution,Luo2014}, and increase levels of activity~\cite{parida2014effect}. The maximum brood sizes for N2 worms in laboratory conditions is achieved for temperatures slightly above $18^{\circ}$C~\cite{begasse2015temperature,gouvea2015experience}. Deviations from this temperatures can cause stress and, because of that,} we used temperature to test different stress conditions in worms. In order to do so, we took worms raised at $20^{\circ}$C and put them at a colder temperature ($15^{\circ}$C) and at a higher temperature ($25^{\circ}$C) and recorded their trajectories. {The worms used for the temperature assays were young adults (day 1 of adulthood)}. Then, we evaluated how the diffusion exponent $\gamma$, fractal dimension $d_f$, and fractal correlation exponent $h$ change with temperature (Fig.\ref{fig:4}).

{For the MSD analysis, temperatures different from $20^{\circ}$C seem to introduce additional noise in the trajectories (in the range $10$s $\,<t<100$s), with bigger effects for the lower temperature; Fig.\ref{fig:4}(a). The distribution of MSD exponents $\gamma$ and statistical differences are shown in  Figs.\ref{fig:4}(d) and (e), respectively. We can identify the three diffusion regimes (ballistic, superdiffusion, and saturation regimes) and differences on the diffusion exponents, similarly to the results for aging; Fig.\ref{fig:4}(f).}

{The fractal dimension $d_f$ for movement behavior decreases as temperature increases; Fig.\ref{fig:4}(b). The correlation exponent $h$ for the velocities time series also change with temperature; Fig.\ref{fig:4}(c). Particularly, we can observe a statistically significant increasing in the correlation exponent $h$ of the head speed as temperature increases.  The response to temperature stimulus seems to affect more head speed since head movements are associated with exploration/sensing during foraging~\cite{notex}. The distributions of fractal dimension, and DFA exponents are shown in Fig.\ref{fig:4}(d) and the matrix of $p$-values are shown in Fig.\ref{fig:4}(e).}

{Despite our efforts to keep temperature constant during the experiments, it is not possible remove small fluctuations in the temperature. It is known that spatial gradients of temperature lead to changes in directionality of motion~\cite{Luo2014}. While it is important to systematically investigate the impact of local temperature on the self-stimulus of trajectory, this goes beyond the scopes of this work.}

The motility of healthy {\it C. elegans} displays fractal properties reminiscent of human physiological signals. As for humans~\cite{Ivanov1999,Goldberger2000,Goldberger2002,Hausdorff2007}, we find statistically significant differences in the fractal behavior of the motility of \textit{C. elegans} for different ages and stress levels. Although, the use of \textit{C. elegans} is already pervasive in biological studies of aging, our results suggest that the similarity to the human aging process is deeper than previously thought; but it extends to subtler perturbations and subtler phenotypes. We believe that \textit{C. elegans} can be used to study how fractal dynamics are created by the regulatory processes of physiological systems and provide insights into the fundamental processes required to maintain a healthy physiology in the face of aging and stress. 

This work has been supported by the agency Coordena\c{c}\~ao de Aperfei\c{c}oamento de Pessoal de N\'ivel Superior (CAPES) under grant 99999.006842/2015-01.


\begin{thebibliography}{50}
\bibitem [{not()}]{notex}%
  \BibitemOpen
  \href@noop {} {}\bibinfo {note} {The temperature does not change the
  properties of the underlying environment. The changes in \textit{C. elegans}
  movement are because of the high sensitivity to temperature that they
  have.}\BibitemShut {Stop}%
\end{thebibliography}
\end{document}